# Potential of a Superconducting Photon Counter for Heterodyne Detection at Telecommunication Wavelength

**M. Shcherbatenko,**[1,2,*] **Y. Lobanov,**[1,2] **A. Semenov,**[1,2] **V. Kovalyuk,**[1] **A. Korneev,**[1,2] **R. Ozhegov,**[1] **A. Kazakov,**[1] **B.M. Voronov,**[1] and **G.N. Goltsman**[1,3]

[1]*Moscow State Pedagogical University, Moscow, Russia*
[2]*Moscow Institute of Physics and Technology, Dolgoprudny, Russia*
[3]*National Research University Higher School of Economics, Moscow, Russia*
*[\*scherbatenko@rplab.ru](mailto:scherbatenko@rplab.ru)*

**Abstract:** Here, we report on successful operation of a NbN thin film superconducting nanowire single-photon detector (SNSPD) in a coherent mode (as a mixer) at the telecommunication wavelength of 1550 nm. Providing the Local Oscillator power of the order of a few picowatts, we were practically able to reach the quantum noise limited sensitivity. The intermediate frequency gain bandwidth was limited by the spectral band of single-photon response pulse of the detector, which is proportional to the detector size. We observed gain bandwidth of 65 MHz and 140 MHz for $7\times7$ $\mu m^2$ and $3\times3$ $\mu m^2$ devices respectively. Tiny amount of the required Local Oscillator power and wide gain and noise bandwidths along with the needless of any Low Noise Amplification opens possibility for a photon counting heterodyne-born megapixel array development.

## 1. Introduction

Using the heterodyne technique, one could achieve a fine frequency resolution limited by the stability and linewidth of the Local Oscillator (LO). Such a technique was originally developed for radio-waves and later on spread into high frequency radio-astronomy. Simultaneously, its application has been promoted for optics [1], with the first experimental proof demonstrated in the before-laser age [2]. With invention of the laser, heterodyne analysis of the signals have been widely exploited in optics, including visible and IR waves, in numerous experiments [3-5], and applications in spectroscopy [6,7], space research [8-10], LIDAR [11,12], optics communication systems [13,14], optical tomography [15-17]. With aim to develop a sensitive tool for the astrophysical observations, heterodyne detection was demonstrated with the use of a p-i-n diode at the telecommunication wavelength [18]. The best sensitivity was achieved in expense of enlarged integration time and quite narrow IF bandwidth which was compensated by the LO sweep. However, a diode–based heterodyne receivers require a significant amount of the LO power [19,20]. Another approach was presented in [24], where a superconducting Hot-Electron-Bolometer mixer made from a thin superconducting film of NbN was used for coherent detection of near-IR radiation. Integration of NbN detector with a plasmonic nano-antennae [21,22] allowed to decrease significantly the LO power preserving a relatively large IF bandwidth which is determined by the hot-electrons relaxation time in the film [23]. However, the LO power required was still quite noticeable (a few tens of microwatts) and sensitivity was strongly dependent on the radiation absorption by the superconducting film. This approach required also special care to be taken of the IF signal processing, including the need for a Low Noise Cryogenic Amplifier followed by a room temperature amplification chain.

An alternative detection approach which combines precise timing resolution of heterodyne technique with the sensitivity of the single quanta detectors has been introduced in [24] and reintroduced in [25]. To achieve such a combination, a photon counter is simultaneously irradiated by the LO and a weak signal beam, which are differ in frequency slightly. Due to the interference between the LO and signal waves the power absorbed by the counter oscillates with the intermediate frequency (which is the difference between the LO and signal frequencies). This variation of the power in time produces the variation of photon flux that can be recorded by the photon counter. Afterwards one is able to restore the incoming signal frequency knowing reference laser frequency and retrieving the intermediate frequency from the recorded power variation, quite analogous to a classical mixer operation. Experimental study of the concept was done with use of a Geiger-mode operated avalanche photodiode (APD) array in [25-27]. Usage of the APD array allowed shortening the system dead-time significantly compared to the operation of a single pixel APD. The laser power was split into two beams, one serving as the signal and the other being the LO shifted in frequency by an acousto-optical modulator. This technique was shown to suit LADAR requirements. However, use of an APD for detection yields some disadvantages, namely, high dark count rate, long dead time and afterpulsing, leading to a limited counting rate [28,29].

In contrast, state-of-the-art superconducting nanowire single-photon detectors (SSPDs or SNSPDs) which respond to each incident photon by a short voltage pulse due to destroying of the superconducting state [30-32] outperform their counterparts in terms of sensitivity and counting rate, demonstrate very low dark count rates (about 0.1 $s^{-1}$), and quite short dead time, typically ~5 ns [31,32]. Several SNSPD design modifications were demonstrated to be capable for the photon-number resolving [33, 34]. When integrated with resonating structures enhancing photon absorption [35], SNSPDs demonstrate above 90% detection efficiency [36] making them useful in a number of applications, such as free-space optical communication [37,38], quantum information processing [39-41], quantum key distribution [42], ranging [43], and life sciences [44,45]. Moreover, arranging SNSPDs into an array may lead [46,47] to even larger system count rates achieved and facilitate spatial resolution which would be beneficial for many practical applications. Recently SNSPDs integrated with optical waveguides [48,49] on a single chip were demonstrated providing a possible solution for on-chip quantum information processing and the quantum computer implementation [50,51].

In this paper we present implementation of the photon-counting heterodyne detection with use of a single SNSPD coupled to a single-mode optical fiber.

## 2. Experiment

For our work, we have exploited a typical meander-shaped SNSPD patterned by e-beam lithography on a 4 nm thick NbN film, with the latter being deposited by reactive DC magnetron sputtering on top of a high resistivity silicon substrate [52]. The nanowire width was 80 nm and spacing between the nanowires was 120 nm. Fig. 1 depicts our experimental setup. The device was installed on top of a single mode optical fiber and a double-wall insert with vacuum insulation was used to bring the bath temperature down to 1.7 K, facilitating the filtering effect of the cooled single mode optical fibers [53]. One laser serves as the LO irradiating at 1550 nm wavelength, and the other one plays a role of the signal. Each laser is a fiber-coupled single mode distributed feedback laser packaged in a standard butterfly case with a thermo-electric cooler allowing for precise laser chip temperature control, allowing for smooth tuning of the LO wavelength. The width of the laser emission line is ~2 MHz, measured at FWHM.

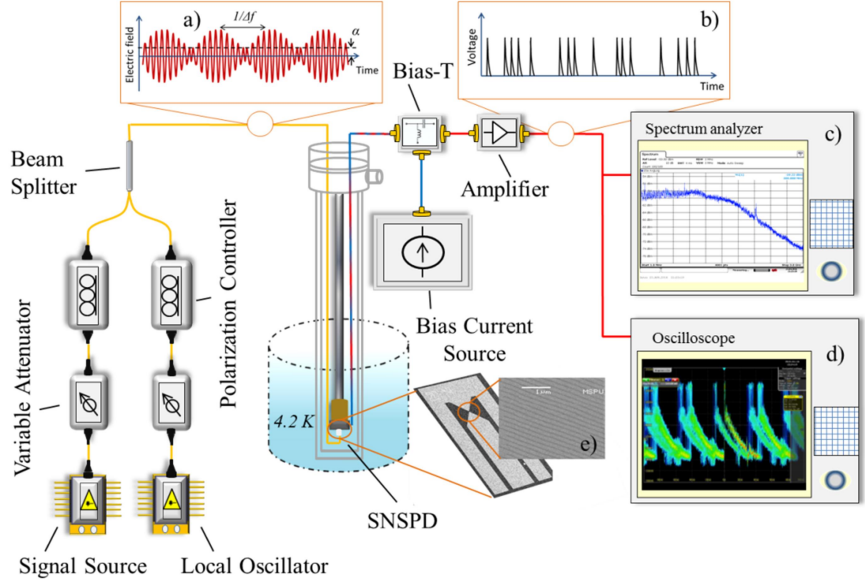

Fig. 1. A schematic view of the experimental setup (see the details in the text). Insets: (a) interference of the LO and signal electric fields incident on the detector, (b) schematic representation of distribution of pulses with time which represent the electric field beating, (c) IF power spectrum in 1 MHz to 3 GHz window due to direct response of the SNSPD to incident radiation of a light source, (d) a train of pulses registered by the oscilloscope, (e) SEM imaging of the SNSPD chip and its central part – meandered NbN film.

Mechanically controlled attenuators and fiber-optics polarization controllers are used to adjust laser power and polarization, respectively. With the use of a fiber-based beam splitter the wave fronts are combined and directed to the SNSPD. The SNSPD is biased by an under-critical current through a bias-tee. In our experiment we used bias current of 24 µA providing quantum efficiency of ~8 %. An absorbed photon destroys superconductivity for a short time giving the rise to resistance and leading to a voltage pulse which can be further amplified and recorded by oscilloscope or any other electronics. When the SNSPD is illuminated by both LO and signal simultaneously (which are set with a slight difference in their wavelength) the waves interfere and the electric field oscillates (as schematically shown in the inset (a) in Fig. 1) at the frequency which is the difference between the LO and signal frequencies. This frequency is traditionally called the "intermediate frequency" (IF). Photon count rate is proportional to the incident power and thus also oscillates with the intermediate frequency resulting in a train of pulses with varied density of the pulses, as depicted schematically in the inset (b) in Fig. 1. The voltage pulses from the SNSPD produced in response to the photon absorption are amplified by the room temperature amplification chain of two amplifiers with total gain of 48 dB and bandwidth of 2 GHz, and the IF signal is further fed in to either an RF signal analyzer or digital oscilloscope.

### 3. Analysis

In principle, one can analyze time-resolved photon statistics to obtain the intermediate frequency $f_{IF}$ and the power of the down-converted signal $P_S$. Theoretical basis of the concept with a rigorous statistical analysis was disclosed in details in [54]. As beatings emerge due to interference between the two light beams of the two sources, and if the field can be treated classically, the power irradiating the detector at a specific moment of time $t$ is:

$$P(t) = (P_S + P_{LO})(1 + \alpha \cos(2\pi f_{IF} t + \varphi)) \qquad (1)$$

Here, the coefficient $\alpha \equiv \frac{2\sqrt{P_S P_{LO}}}{P_S + P_{LO}}$ determines the relative depth of the modulation. Finite spectral width of the beatings can be accounted for by stochastic slow time variations of the phase $\varphi$.

In case when the detector resolves single photons, the field should be considered as quantized one. In this case the power given by (1) has the meaning of the average over ensemble of measurements and determines probability to detect a photon per unit time:

$$\frac{d}{dt} p(t) = \eta \frac{P_S + P_{LO}}{hf} (1 + \alpha \cos(2\pi f_{IF} t + \varphi)) \qquad (2)$$

with $\eta$ being the detection efficiency. To extract power of the signal and its spectrum from the sequence of the pulses, one can use Fourier-analysis: the power spectrum density (PSD) of the Fourier-transformed voltage trace from the detector containing many photon counts is

$$PSD(f) = \varepsilon(f) r_S r_{LO} s_{IF}(f) + [\varepsilon(f)(r_S + r_{LO} + r_D) + S_{el}(f)] \equiv PSD_S(f) + PSD_N(f), \qquad (3)$$

where $s_{IF}(f)$ is the power spectrum of the beatings (having maximum at $f_{IF}$ and normalized such as $\int s_{IF}(f) df = 1$, $\varepsilon(f)$ is the energy spectrum density of single voltage pulse from the detector, $r_S$, $r_{LO}$ are the count rates which were caused by the signal and local oscillator lasers independently, i.e. without interference, $r_D$ denotes the rate of dark counts of the detector, and the term $S_{el}(f)$ is introduced to account for spectral density of additional electrical noise (added by amplifiers and the IF chain), which plays a role if the IF signal is processed by the spectrum analyzer (but would be absent if only time moments of the photon counts are recorded and the Fourier analysis is applied). Energy spectrum density of the single voltage pulse from the detector can be expressed through the Fourier transform of the single voltage pulse $v(f)$ and the input impedance of the spectrum analyzer $R$ as $\varepsilon(f) = |v(f)|^2 / R$. The signal contribution is presented by the first term in (3), and the other terms proportional to $r_{S,LO,D}$ should be considered as a statistical noise, related to the fluctuation of photon and dark counts. According to (3), gain bandwidth is limited by the energy spectrum of the pulse $\varepsilon(f)$ and we shall proof this experimentally below.

One can use the RF spectrum analyzer to perform Fourier transformation in hardware. Alternatively, only the moments of time when the counts are registered can be recorded and then processed in line with the schematic sketched above. In this case the pulse spectrum and the electrical noise drop out, but the post-processing is required. Both approaches should produce same result. We tested both methods in our experiments and the results are compared in Fig 2.

First we used the power spectrum analyzer to obtain the spectrum shown in Fig. 2a. When SNSPD is irradiated by either the LO or signal laser, output of the spectrum analyzer is the spectrum of a single SNSPD pulse only with its width defined by the duration of the pulse and electronics bandwidth. When both LO and signal lasers are in operation one can observe a peak (red curve marked "LO + signal" in Fig 2a) in the spectrum which is located exactly at the frequency which is the difference between the LO and signal laser frequencies. In our case the LO and signal are shifted in frequency by 10 MHz. The spectrum of a single source represented by a green curve and marked "1 source" in Fig 2a acts as a noise floor for the signal.

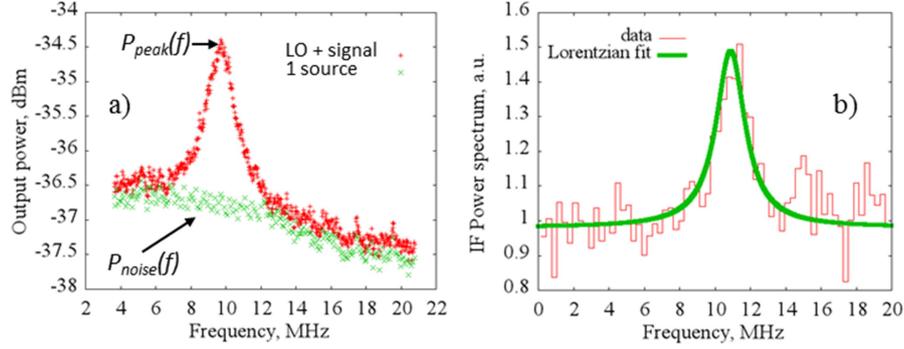

Fig 2. (a) SNSPD output signal processed with the RF spectrum analyzer with resolution bandwidth (RBW) set to 300 kHz. Red curve is obtained when SNSPD is illuminated by both LO and signal lasers. The signal at the IF is ~2 MHz wide peak (marked $P_{peak}(f)$). The noise floor (green curve marked as $P_{noise}(f)$) is essentially the spectrum of the single-photon response pulse of the SNSPD. (b) The same result is obtained by mathematical Fourier-analysis of the 1-ms-long trace of the SNSPD pulses recorded with the digital oscilloscope. Solid (green) curve is the fit by the Cauchy-Lorentzian distribution formula.

Then, we used a 1 Gs/sec digital oscilloscope to acquire an oscillogram of 1 ms long with time resolution of 1 ns. With mathematical post-processing, we determined moments of time $t_k$ at which the photo-counts occurred, calculated Fourier transform as the sum of exponents with arguments proportional to those times $\sum_k exp(2\pi i f t_k)$, and then obtained power spectrum shown by red line in Fig. 2b which reproduces reasonably well the analog spectrum obtained with the RF spectrum analyzer. The green solid line is the fit by the Cauchy-Lorentzian distribution formula: $f(x,x_0,\gamma) = [1/(\pi \gamma)] \cdot [\gamma^2/[(x-x_0)^2 + \gamma^2]]$ with $x_0$ the peak frequency, $2\gamma \approx 2$ MHz the peak width. One can see that both methods produce similar results, meanwhile mathematical processing of the digital data from the oscilloscope gives the result with already eliminated spectrum of the voltage pulse. Below we present the results obtained by analog processing with use of the RF spectrum analyzer.

To characterize signal and noise properties of our receiver, we measured dependence of output *PSD* vs signal power at $f=f_{IF}$=15 MHz and different LO powers. Fig. 3 presents an example of such dependence measured at 15.6 pW LO power. Red points correspond to the *PSD* at $f=f_{IF}$, whereas blue points are PSD at the same frequency but with changed $f_{IF}$, such as the peak is shifted away, i.e., blue points correspond to $PSD_N$ and red ones correspond to $PSD_{peak}=PSD_S(f_{IF})+PSD_N(f_{IF})$. Hence, to calculate $PSD_S$, one has to subtract $PSD_N$ from $PSD_{peak}$.

To estimate absolute gain of our detector, one needs to know signal power and gain of the IF chain. As one can extract from data show in Fig. 3, at signal laser power of 1 pW the output power in the band RBW=300 kHz is $2\times10^{-8}$ W, and the total power, in the band $\Delta f$ = 2 MHz, is $P_S$=1.3×10$^{-8}$ W. Accounting for gain of the IF chain (48 dB) and DE=0.08, one finds that our SNSPD exhibits intrinsic gain of ~ 10 dB, which we relate to the operation principles of the SNSPD: as a photon is absorbed, SNSPD generates a voltage pulse whose amplitude is determined by the Joule power released while the SNSPD is in the resistive state. This is a kind of internal amplification mechanism allowing one to increase significantly the gain which essentially results in the system sensitivity, eliminating the need for a low-noise amplification stage in the IF chain.

As a measure of noise, one can use ratio of PSD of the output signal ($PSD_S$) at the frequency $f_{IF}$ (corresponding to its maximum) to PSD of the noise ($PSD_N$). According to formula (3), this ratio is

$$\frac{PSD_S(f)}{PSD_N(f)} = \frac{r_S r_{LO} s_{IF}(f)}{r_S + r_{LO} + r_D + S_{el}(f)/\varepsilon(f)} < r_S s_{IF}(f) \equiv \frac{r_S}{\Delta f} \frac{s_{IF}(f)}{s_{IF}(f_{IF})}. \quad (4)$$

In the latter equality, we used the definition of $\Delta f$, $\int s_{IF}(f)df = \Delta f s_{IF}(f_{IF})$, and the normalization condition $\int s_{IF}(f)df = 1$. Inequality in RHS of the formula sets the so-called quantum limit of noise: to reach $PSD_S/PSD_N = 1$, one requires at least one registered photon from the signal source per time $1/\Delta f$. Actually, the ratio of PSD (formula 4) defines signal-to-noise ratio (SNR) if only PSD in the neighboring spectral bins can vary independently, namely if one has no *a-priori* information about the $s_{IF}(f)$ or cannot resolve the spectral width of the beatings $\Delta f$. In contrary, if, as in our experiment, one knows that there is a single line, the spectral resolution of the spectrum analyzer RBW<$\Delta f$, and the goal is just to measure the signal power on input (but not its distribution over the spectrum), the SNR should rather be defined as

$$SNR = \frac{P_S}{P_N} = \frac{\int PSD_S(f)df}{RBW \times PSD_N(f_{IF})} = \frac{PSD_S(f_{IF})}{PSD_N(f_{IF})} \frac{\Delta f}{RBW} \tag{5}$$

Substituting here the formula (4) one comes to

$$SNR = \frac{r_S r_{LO}}{r_S + r_{LO} + r_D + S_{el}(f_{IF})/\varepsilon(f_{IF})} \frac{1}{RBW} < \frac{r_S}{RBW} \tag{6}$$

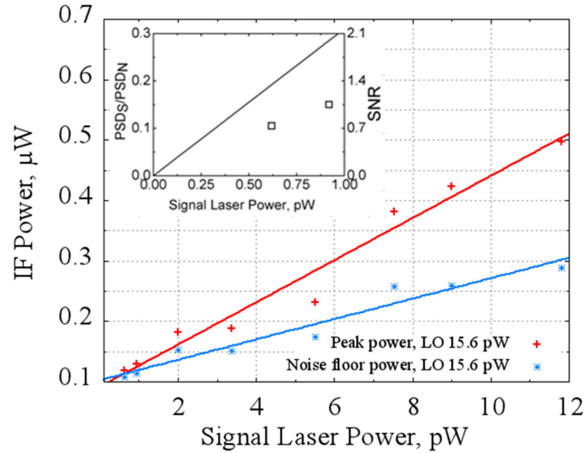

Fig. 3. IF power vs input signal power for 15.6 pW LO power measured at RBW 300 kHz. Lines are guides for an eye. Inset shows the ratio of PSD of output signal to PSD of noise (left vertical axis) and corresponding SNR (right axis) for minimal levels of signal power. Line is the quantum limit of noise for DE=0.08.

The inset in Fig. 3 demonstrates near quantum limited level of noise achieved at the minimum power of signal on the input $P_{Sin}$ (below 1 pW). The quantum limit of noise given by inequalities in the RHS of (4) or (5) is represented by straight line. Rate of counts, figuring there, was expressed through $P_{Sin}$ as $r_S = DE \times P_{Sin}/h\nu$, where $h\nu$ is the photon energy.

Another important characteristic of a practical heterodyne receiver is the intermediate frequency (IF) bandwidth, which determines maximum separation between the LO and incident signal in frequency allowing yet for an efficient signal acquisition. We have measured both gain and SNR intermediate frequency bandwidths of the SNSPD based heterodyne receiver. By tuning the LO wavelength (while the amplitude and frequency of the signal source remained unchanged) we measured the $PSD_S(f_{IF})$. By definition, this quantity is proportional to frequency-dependent gain. We defined Gain Bandwidth (GBW) as the frequency $f_{3dB}$ where the Gain is reduced by a factor of 2.

The measurement results are summarized in Fig. 4 for SNSPDs of two sizes: 7×7 μm² and 3×3 μm². Smaller detector has shorter pulse duration due to a lower kinetic inductance [55]. For 7×7 μm² detector GBW is measured to be 65 MHz (blue crosses in Fig 4a). To proof that the GBW is limited by the spectrum width of the single-photon pulse of the SNSPD we performed Fourier transform of 7×7 μm² SNSPD single-photon pulse (shown in the inset in Fig. 4a). The calculated spectrum is plotted in Fig. 4a as open magenta squares. The 3×3 μm² detector has the GBW of 140 MHz due to its shorter photo-response pulse duration.

Similarly, the Signal-to-Noise Bandwidth (SNR-BW) is defined as the intermediate frequency at which the SNR is reduced by a factor of 2, with the $SNR(f)$ given by (5). Fig. 4b shows the measured SNR-bandwidth of ~1400 MHz and 1200 MHz for the same pair of detectors. We directly compare measured SNR to (6). To perform the comparison, one has to know $r_S$, $r_{LO}$, $S_{el}(f)$ and $\varepsilon(f)$. $S_{el}(f)$ was obtained from distinct set of measurements of PSD on the output of spectral analyzer at different input power of single LO laser on input, including zero power. It was found to be nearly independent on frequency in the range of interest, from 0 to 1500 MHz, and equal to $4.5\times10^{-16}$ W/Hz. $\varepsilon(f)$ was obtained from Fourier-transform of digitized voltage pulse of the SNSPD, its dependence on frequency is presented in Fig. 3a and its absolute value at f=50 MHz was $2\times10^{-22}$ J/Hz. Then we calculated $r_S+r_{LO}$, dividing measured $PSD_N(f)$ at 50 MHz, where it was several order of magnitude greater than $S_{el}(f)$ (thus the latter can be neglected) to $\varepsilon(f)$. Analogously, we calculated the product $r_S r_{LO}$ by dividing measured $PSD_S(f)$ to $(\varepsilon(f)/\Delta f)$. We repeated this calculation for several greater values of frequency, taken from the range where frequency dependence of $PSD_{N,S}(f)$ coincided with $\varepsilon(f)$ (i.e. no nonlinear distortions of the pulse occurred) and at the same time $S_{el}(f) \ll PSD_N(f)$, and proved independence of the extracted rates on the frequency. The resulted plots are presented in Fig. 4b as solid curves.

Hence, we demonstrated that measured values are dominated by our measurement setup system noise. We believe that ultimate SNR BW of SNSPD based receiver should be related to the detector jitter, which is less than 20 ps for this type of detector [56]. With increase of the IF, the time uncertainty $d\tau$ of the pulse appearance in regard to the photon absorption, i.e. jitter, becomes comparable with the periodicity of the beatings ($2\pi/f_{IF}$) leading to their averaged amplitude decrease at IF approaching $2\pi/d\tau$ and its complete vanishing at IF $\gg 2\pi/d\tau$.

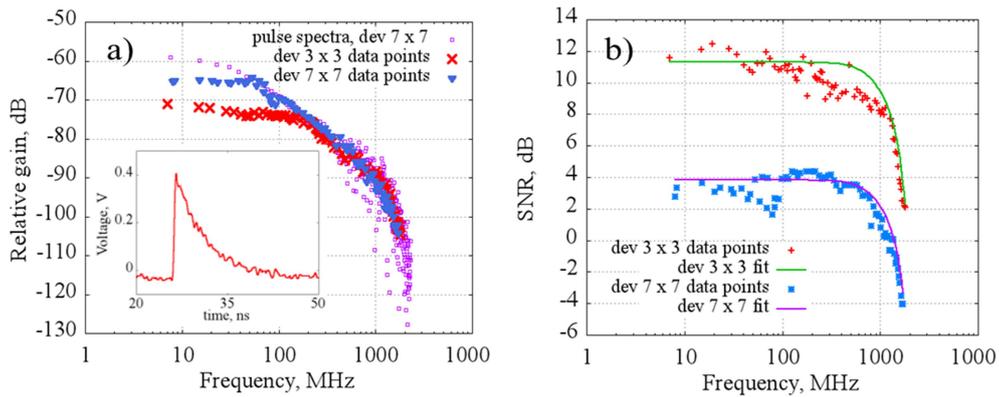

Fig. 4. (a) Gain bandwidth (GBW) measured for SNSPDs of two sizes: 7×7 μm² (blue crosses) and #2: 3×3 μm² (red triangles), which differ in photo-response pulse duration. The GBW is limited by the duration of the single-photon response pulse: magenta squares is the Fourier transform of the single-photon response of 7×7 μm² SNSPD shown in the inset. (b) Signal-to-noise ratio (SNR) bandwidth for 3×3 μm² SNSPD (red) and 7×7 μm² SNSPD (blue).

## 4. Conclusion

To conclude, we have demonstrated that NbN thin film based superconducting nanowire single photon detector can be operated in the heterodyne regime. In this regime it requires local oscillator power of order of a few picowatts to resolve subpicowatt optical signals. We demonstrated 65 MHz gain bandwidth for 7×7 μm$^2$ device limited by single-photon pulse duration, and 140 MHz for 3×3 μm$^2$ device with a shorter response pulse. The SNR bandwidth was ~1400 and ~1200 MHz for 7×7 μm$^2$ and 3×3 μm$^2$ device respectively. Although our detector had a somewhat poor quantum efficiency, it still allowed us to demonstrate the coherent mode sensitivity limited by the quantum noise only. This opens a wide possibility for detector array development. Combining the advantages of heterodyne method and single-photon detection, it is possible to produce power spectral analysis of weak signals with a high spectral resolution and sensitivity at the quantum noise limit.

## Funding

This work was supported by the Ministry of Education and Science of the Russian Federation, contract no. 14.B25.31.0007 and by Russian Foundation for Basic Research, contract no. 16-32-00465.

## Acknowledgments

We would like to thank prof. T.M. Klapwijk for stimulating discussions.